# Photoluminescence and High Temperature Persistent Photoconductivity Experiments' in SnO$_2$ Nanobelts.

E. R. Viana*, J. C. González, G. M. Ribeiro and A. G. de Oliveira.

Departamento de Física, Universidade Federal de Minas Gerais, Av. Pres. Antônio Carlos, 1500, Pampulha, 31270-901, Belo Horizonte/MG – Brazil.

Corresponding author, e-mail: emilsonfisica@ufmg.br, emilson.fisica@gmail.com

**Abstract**

The Persistent Photoconductivity (PPC) effect was studied in individual tin oxide (SnO$_2$) nanobelts as a function of temperature, in air, helium, and vacuum atmospheres, and low temperature Photoluminescence measurements were carried out to study the optical transitions and to determine of the acceptor/donors levels and their best representation inside the band gap. Under ultraviolet (UV) illumination and at temperatures in the range of 200 to 400K we observed a fast and strong enhancement of the photoconductivity, and the maximum value of the photocurrent induced increases as the temperature or the oxygen concentration decreases. By turning off the UV illumination the induced photocurrent decays with lifetimes up to several hours. The photoconductivity and the PPC results were explained by adsorption and desorption of molecular oxygen at the surface of the SnO$_2$ nanobelts. Based on the temperature dependence of the PPC decay an activation energy of 230 meV was found, which corresponds to the energy necessary for thermal ionization of free holes from acceptor levels to the valence band, in agreement with the photoluminescence results presented. The molecular-oxygen recombination with holes is the origin of the PPC effect in metal oxide semiconductors, so that, the PPC effect is not related to the oxygen vacancies, as commonly presented in the literature.

_________________________________________________________________________________________

## 1. Introduction

One dimensional nanostructures have stimulated significant contributions in scientific research in recent years. Because of its small diameter, d ≤ 100nm, the surface effects and the quantum confinement can influence their optical [1-5] and electrical [6,7] properties which are important for the functionality and performance of nanodevices [8]. In particular, tin oxide (SnO$_2$) nanowires and nanobelts[9] have large potential for applications as gas sensors [10, 11] and as ultraviolet light detectors [12], because of their inherent high surface-to-volume ratio properties. However, a deeper understanding of the optoelectronic properties of the SnO$_2$ nanostructures is still needed.

In this work, the photoconductivity (PC) and the photoluminescence (PL) of SnO$_2$ nanobelts was studied. PC measurements were carried out as a function of temperature, and at different atmospheres, namely, in air, helium and vacuum. Low temperature PL experiments were also carried out, to study the optical transitions that will be used to the determination of the acceptor/donor energy levels involved, and to support the photoconductivity studies.

When exposed to UV illumination the SnO$_2$ devices show a fast and strong enhancement of the photoinduced current, and the maximum value obtained increases with a temperature decreasing. On the other hand, when the UV illumination is turned off the current decays for hours, even at temperatures as high as 400 K which characterizes the phenomena named Persistent Photoconductivity (PPC).

The photoconductivity and the PPC results were explained in terms of oxygen adsorption and desorption at the surface of the SnO$_2$ nanobelts that promote the creation of a band bending, a spatial charge separation and a decrease in the free carriers' recombination time. The temperature dependence of PPC was found to exhibit an activation energy of 230 meV, which corresponds to the energy necessary for thermal ionization of free holes from acceptor levels to the valence band, level determined from the PL experiment. So, holes from acceptors levels are the responsible for the PPC effect in SnO$_2$, when we are using a light-source with energy below band-gap.

Some authors have reported and proposed models in order to explain the PPC effect in metal-oxide semiconductor nanowires, such as in ZnO [3,4,29] and in WO$_3$ [14]. However, there is still no conclusive proposal about the origin of PPC on oxide semiconductors [3,4,29], that is subject of the present work.

## 2. SnO$_2$ nanobelts: synthesis and characterization

The SnO$_2$ nanobelts used in this work, were synthesized by using a gold-catalyst-assisted VLS method [13]. A suitable amount of pure Sn powder (1g,

99.99% purity) was placed on top of a Si:SiO$_2$ substrate, previously coated with a thin Au film. The substrate was placed in a horizontal quartz tube furnace that was then heated up to 800 °C for 2 hours, and after that let to cool down naturally. SnO$_2$ nanobelts were found on the substrate surface in a cotton-wool-like form.

TEM examinations, see **fig.1**, show that nanobelts present a prismatic cross-section and are covered by a 10-20nm thick amorphous tin oxide layer. SEM examinations, see **inset fig.1**, reveal a large quantity of nanobelts with typical widths in the range of 50 to 500 nm, and lengths from 5 to 50 μm. XRD measurements, not showed, confirms the tetragonal rutile crystal structure of the nanobelts with lattice constants of a = b = 0.473nm and c = 0.318nm [15].

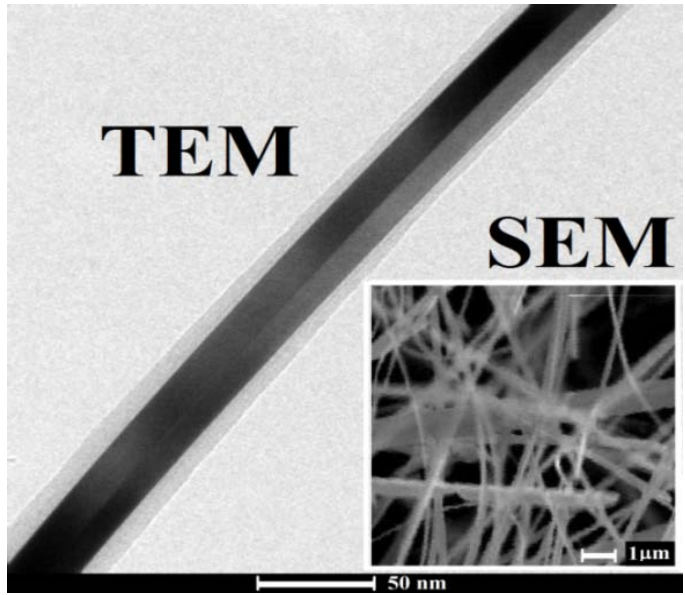

**Figure 1.** (a) The TEM image of a SnO$_2$ nanobelt covered by a thick amorphous layer. A SEM image of several SnO$_2$ nanobelts is shown in the inset.

### 3. Photoluminescence Experiments

Low temperature photoluminescence experiments were carried out to study the optical transitions in the SnO$_2$ nanobelts, and to support the photoconductivity studies that will be presented.

A 325 nm line of a HeCd laser, *Kimmon instruments®*, was used to perform the photo-luminescence (PL) experiments in a bundle of SnO$_2$ nanobelts. The PL emission was analyzed with *a PI-Acton® SpectraPro-2558i* spectrometer equipped with an UV-enhanced liquid nitrogen cooled Si charge couple device (CCD) detector. PL was measured at 4 K in a cold-finger *Janis®* ST-100 cryostat, with standard UV grade optical view ports.

**Figure 2** shows the PL spectrum of the SnO$_2$ nanobelts at 4K. In the sub-UV region (infrared (IR), visible (VI)) of the spectrum **fig.2.(a)**, six peaks were observed at: 2.57, 2.07, 1.47, 1.40, 1.32 and 0.90 eV. The strong and broad orange (2.07 eV) and green (2.57 eV) gap luminescence is assigned to defect states due to oxygen vacancies in the tin oxide, as those recently published by A. Kar et. al [30,31]. Oxygen vacancies in three charge states have been noted in SnO$_2$, neutral $V_\varphi$, single $V_o^+$ and double ionized $V_o^{++}$. The peak 2.07 eV is believed to be the $V_o^+$ state, and the peak at 2.57 eV is belived to be assigned to isolated $V_o^+$ centers, which lies in higher energy than the complex $V_o^+$ center band at 2.07 eV [30,31]. The IR-peaks, 0.90, 1.32, 1.40 and 1.47 eV, are assumed to be transitions from trapped electrons, forming donor states ($E_{D1}$, $E_{D2}$ and $E_{D3}$), to the mid-gap oxygen vacancies states, ($V_o^+$) and ($V_o^+$)$_{iso}$, that were defined by the broad and strong orange/green peaks.

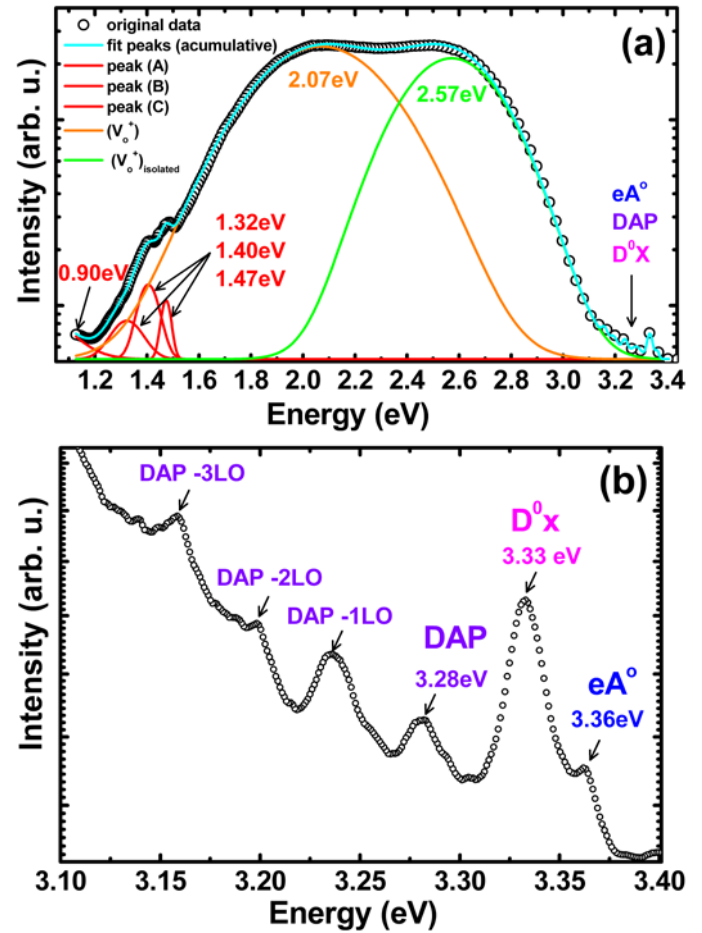

**Figure 2.** (Color Online) **(a)** The photo-luminescence (PL) spectrum (IR-VI and UV region) of the investigate SnO$_2$ nanobelt sample at 4K, showing a strong and broad orange and green luminescence, at 2.07 and 2.57 eV, respectively. **(b)** UV PL-spectra of SnO$_2$ nanobelts taken at 4K, showing

the eA°, D⁰x and DAP transitions, and LO replicas of the DAP transition.

In the UV region of the spectrum **fig.2.(b)**, six peaks were observed at: 3.16, 3.20, 3.24, 3.28, 3.33 and 3.36 eV. The peaks at 3.36 and 3.33 eV are assigned to the recombination of electrons from the conduction band to acceptor levels (eA°) and to excitons bound to neutral ($D^0X$)[16-18], respectively. The peaks at 3.28, 3.24, 3.20 and 3.16 eV are associated with donor-acceptor pairs (DAP) recombination and the DAP-1 to DAP-3 are the longitudinal phonon (LO) replicas. All PL peaks of the SnO₂ nanobelts observed in **fig.2** are in agreement with previous reports [16-18, 30-31].

Considering the accepted value of 3.60 eV for the energy-gap of the SnO₂ nanobelts [19,20], the ionization energy of the acceptor and the donor level involved in the optical transitions eA° and DAP could also be determined to be, $E_A$= 240 meV and $E_{D1}$= 80 meV. The donor level $E_{D1}$ could also be defined by the IR-UV peaks, 2.07 and 1.47 eV, since both are broad peaks, the value of $E_{D1}$ obtained was 60 meV, so in order to corroborate with the other transition the donor state were defined to be in the range of $E_{D1}$ = 60-80 meV.

By transitions between mid-gap $V_o^+$ states, another two donors levels could also be determined as $E_{D2}$ = 130 meV and $E_{D3}$ = 210 meV. Finally the schematic energy diagram of these transitions is shown in **fig.3,** such that down-arrows are the optical transitions, $V_o^+$ and $(V_o^+)_{iso}$ are the oxygen-vacancies states, and the double-arrows are the donor/acceptor levels.

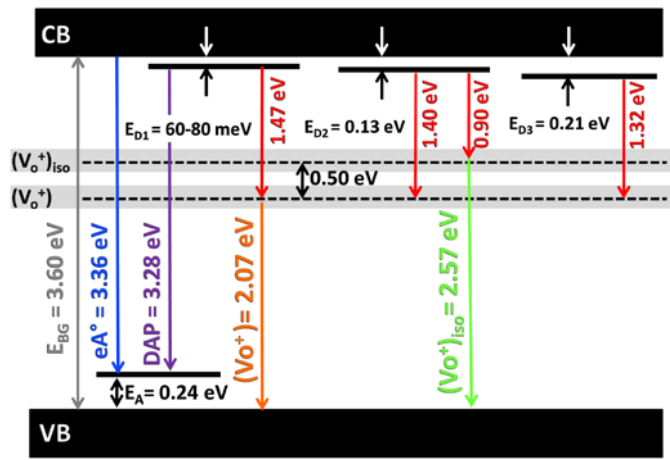

**Figure 3.** (Color Online) Schematic energy band diagram and the optical transitions in the SnO₂ nanobelts.

## 4. Device fabrication

In order to measure the photoconductivity (PC) of a single SnO₂ nanobelt, devices with two contacts were fabricated by photolithography using a LaserWriter® (UV-Laser, λ=405 nm), and standard lift-off process. Electrodes, 2-3 μm apart, consisting of 10 nm of Cr followed by 150 nm of Au were defined on top of a 300 nm thick SiO₂ deposited on a highly-doped p-type Si substrate. The SnO₂ nanobelts are surrounded by an amorphous high resistivity oxide layer, so a H₂-plasma treatment was carried out to improve the quality of the contacts. A schematic of the device is shown in **fig.4**.

## 5. Photoconductivity Measurements

For the PC measurements the SnO₂ device was placed in a cold-finger *Oxford*® cryostat in He-7 atmosphere, with a precise temperature controller, and a *Keithley*® 237, source-meter was used to measure the current $I_{ds}$ passing through the nanobelts as a function of time t, with a constant applied voltage $V_{ds}$. An UV light emitting diode (LED) emitting at 403 nm (3.08 eV) was placed near the sample and used as excitation source in the PC experiments.

**Figure 5** shows the PC measurement results at temperatures of 300, 350 and 400 K, acquired with applied voltage $V_{ds}$ = 100mV and a UV-light intensity of 20.56 mW/cm². Upon exposure to UV illumination, a fast increase in the photocurrent $I_n$ was observed. We can also observe that the generation rate of $I_n$ do not changed significantly with temperature, but the maximum induced photocurrent $(I_n)_{máx}$ increases with a temperature decreasing.

When the UV-light was turned off, the current slowly decays as shown in **fig.5**, and continues decaying for hours and even at temperatures as high as 400 K. This is characteristic of the phenomena namely Persistent Photoconductivity (PPC).

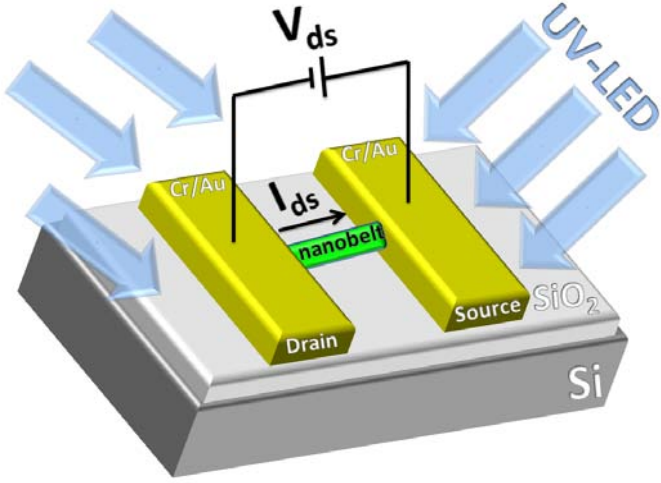

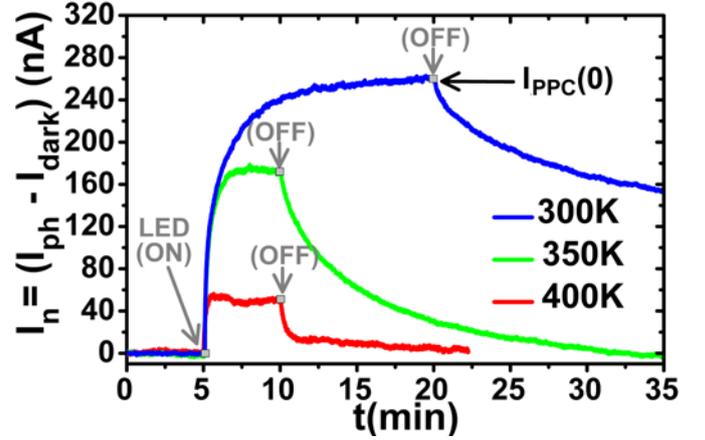

**Figure 4.** (Color Online) Schematic two-contacts device used in this work to measure the photoconductivity of individual nanowires.

**Figure 5.** (Color Online) The time-response of the induced photocurrent $I_n$ for the SnO$_2$ nanobelts, as a function of temperature.

The decay of $I_n(t)$, the photocurrent as a function of time, and normalized to $I_{PPC}(0)$ as show in **fig.6** can be well described by the stretched-exponential function[21]:

$$I_{PPC}(t) = I_{PPC}(0).exp(-(t/\tau)^\beta) \quad (eq.(1))$$

where $\tau$ and $\beta$ represents, the lifetime of the induced photocarriers and the decay-exponent, respectively. Both parameters, $\tau$ and $\beta$, were determined by fitting the $I_{PPC}$ curves with **eq.(1)**, for each temperature measured, as show in **fig.6**. This stretched-exponential function is usually used in the barrier model [21-25] to explain the PPC effect caused by traps such as AX centers in GaN [23], and DX [24] and EL2[25] centers in AlGaAs.

In processes in which trapped electrons or holes are being thermally freed from traps to the conduction or to the valence band, the temperature dependence of the photocarriers lifetime can be expressed as [26]:

$$\left(\frac{1}{\tau}\right) = N_c \delta_n \exp\left(\frac{-\Delta E_{trap}}{k_B T}\right) \quad (eq.(2))$$

where $N_c$ is the effective density of states in the conduction band, $\delta_n$ is the capture coefficient of the traps for electrons, $\Delta E_{trap}$ is the estimated trap-depth or ionization-energy, and $k_B$ the Boltzmann constant.

**Figure 7** shows the temperature dependence of the parameters $\tau$ and $\beta$ in the range of 200K to 400K. A strong decrease of $\tau$ and a slowly increase of $\beta$ are observed as the temperature increases. The activation energy of $\tau$, $\Delta E_{trapp}$= 230 meV, was determined by fitting the temperature dependence of $\tau$ with **eq. (2)**.

In order to explain the PPC behavior in the SnO$_2$ nanobelt and the high value achieved for $\Delta E_{trapp}$ photocurrent experiments were carried out in different atmospheres: air, helium, and vacuum. The photoresponse were measured at fixed temperature of 300K and with the same UV light intensity define previously.

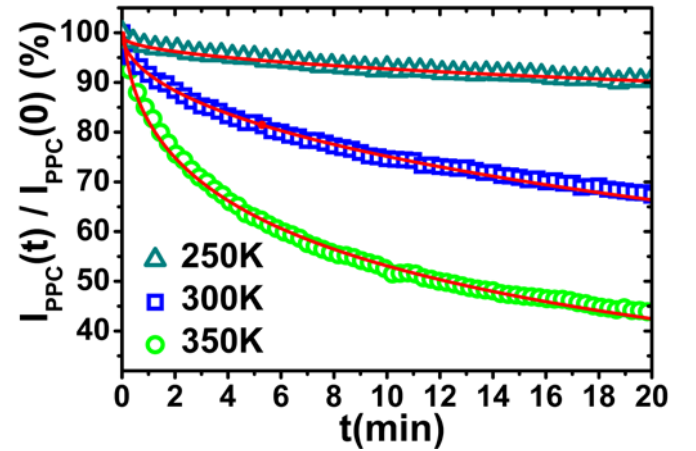

**Figure 6.** (Color Online) Normalized decay photocurrent $I_{PPC}(t)/I_{PPC}(0)$ curves at temperatures 250, 300 and 350K. The solid lines are the respective fittings using Eq. (1).

**Figure 8** shows that the maximum induced photocurrent $(I_n)_{máx}$ is already two times higher in vacuum than in helium, and at least four times higher than in air. Besides, the decay time in vacuum ($\tau_{vacuum}$ = 1.20x10$^5$ s) is almost twenty times longer than in helium ($\tau_{helium}$ = 7.35x10$^3$ s) and one thousand times longer than in air ($\tau_{air}$ = 1.20x10$^2$ s). These results indicate that the presence of oxygen in the atmosphere determine the change of the photoconductivity of the SnO$_2$ nanobelt.

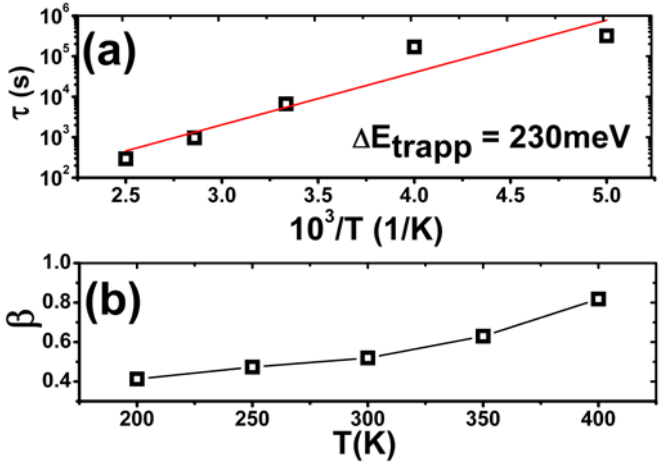

**Figure 7.** Temperature dependence of the PPC lifetime τ (a), and of the decay exponent β (b). The solid line corresponds to the fitting of the experimental data with Eq. (2).

## 5. Discussion

It is believed that the nature of the oxygen surface species depends on the temperature. Oxygen adsorbs on SnO$_2$ surface in a non-dissociatively, i.e. in molecular form, either as neutral O$_2$(ads) or charged O$_2^-$(ads) at low temperatures (T < 500K). At higher temperatures (T > 500K), O$_2$(ads) dissociates into atomic form, as neutral O$_{(ads)}$ or in charged O$^-_{(ads)}$[30].

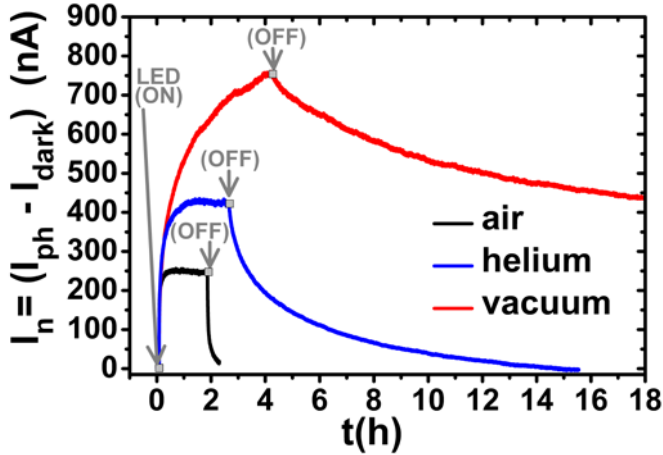

**Figure 8.** (Color Online) The time-response of the induced photocurrent $I_n$ for the SnO$_2$ nanobelts for different atmospheres, air, helium and vacuum, taken at 300K.

The molecular oxygen O$_2$(g) trapp electrons according to:

**Adsorption:** $O_2(g) + e^- \rightarrow O_2^-(ad)$   (eq.(3))

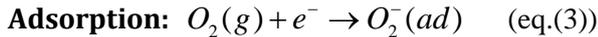

where $O_2(g)$ and $O_2^-(ad)$ indicate oxygen in its free and adsorbed states, respectively.

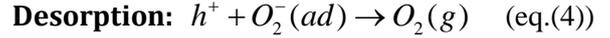

The reverse process, namely the desorption of oxygen from the surface, requires a photogenerated hole $h^+$:

**Desorption:** $h^+ + O_2^-(ad) \rightarrow O_2(g)$   (eq.(4))

Considering these two processes and the schematic energy band obtained from the PL measurements, we want to explain the PPC phenomena in the SnO$_2$ nanobelts.

Once the UV light is turned on, oxygen molecules (that is supposed to be naturally adsorbed at the range of temperatures measured) are desorbed from the surface, and photoexcited holes become available, thereby reducing the surface potential and the depletion width, until a steady state with photocurrent $(I_n)_{máx}$ is reached, as presented in **fig.9**.

As presented in **fig.5** and **fig.8**, oxygen desorbs in vacuum, air or helium (at temperatures between 200-400K) at approximately the same rate. However, the steady-state value, $(I_n)_{máx}$, is different due to the available concentration of molecular oxygen in each case.

As the illumination is turned off and at temperatures near the room temperature, the adsorbed oxygen molecules trap the electrons thermally activated, gradually bend the bands, raising the surface potential barrier and the internal field, as presented in **fig.10**. This reduces the chance of new electrons being trapped at the surface, thereby reducing the rate of oxygen adsorption, which promote a partial separation between the electrons and the photogenerated holes, resulting in an enhancement in their recombination time [3].

This explains the PPC effect and the observed difference between the PPC in air, helium and vacuum. So, the changes in the surface potential and in the depletion width are determined by the interplay between oxygen adsorption (**eq.(3)**) and the net desorption rate (**eq.(4)**), and consequently determining the photoconductivity behavior. But how to explain the temperature dependence?

The activation energy $\Delta E_{trapp}$ associated with the decay time was found to be about 230 meV. This value is very close to the ionization energy of the acceptor level $E_A$= 240 meV found in the PL experiments, as can be seing in **fig.3**. The increase of free holes density due to ionization thermally activated will further decrease the band bending at the surface **fig.10**, which in turn, reduce the spatial separation and recombination time of free electrons and holes when the illumination is turned off, in agreement with our observations.

This way, we understand that the temperature dependence of the PPC decay time is due to the thermal ionization of holes from the acceptor level. The high value of the activation energy obtained explains the observation of PPC even at temperatures as high as 400K. This is the main contribution of the present paper, since many authors have proposed that the mechanisms behind PPC on metal oxide semiconductors are related to oxygen adsorption, desorption, but given no support for the temperature dependence [3,12].

Undopped $SnO_2$ typically shows n-type conductivity. In order to prove that, I-V curves (not showed) as a function of gate bias, and carried out before illumination and also after the illumination is turned off, confirming that the conductivity in our $SnO_2$ nanobelts is n-type. By the results presented in **fig.3**, the n-type behavior of the $SnO_2$ nanobelts at near room temperature can be explained by the shallow donor stated $E_{D1}$= 60-80meV, since part of the carriers can be thermally activated ($E_{D1} \leq 3.k_BT$) from $E_{D1}$ to the conduction band. The n-type behavior may not be explained from the donor state $E_{D2}$, $E_{D3}$ that is too deep to be thermally activated, neither from the oxygen-vacancies surface levels, ($V_o^+$) and ($V_o^+$)$_{iso}$, that are even more deep. So the source of the electrons in the process of oxygen $O_2(g)$ adsorption (**eq.(2)**) are not from the oxygen vacancies.

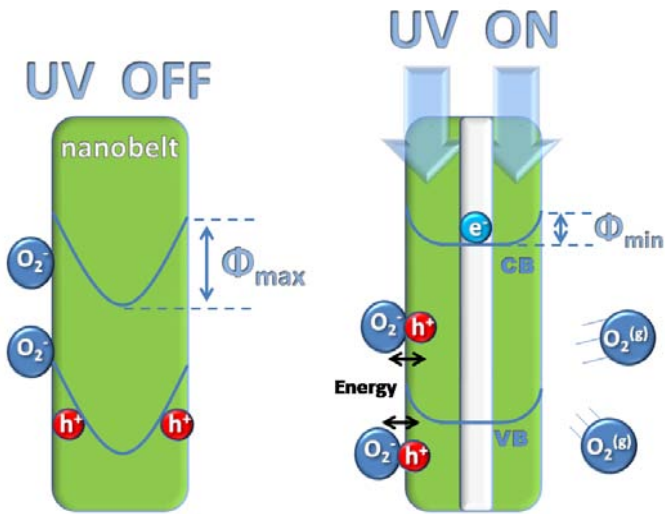

**Figure 9.** (Color Online) UV photogenerated hole promotes desorption of the negatively charged adsorbed oxygen at the surface, reducing the surface potential Φ and the depletion width, increasing the photocurrent.

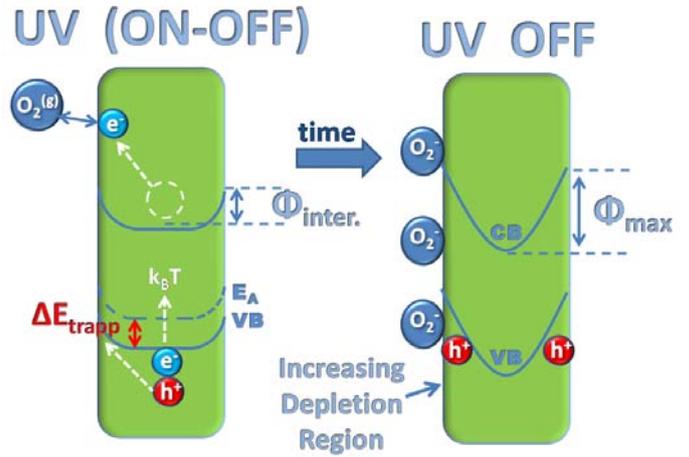

**Figure 10.** (Color Online) Schematic oxygen absorption at the surface process by an electron trapping, increasing the surface potential Φ and the depletion width.

Since an LED emitting at 403 nm (energy of 3.08eV), is used as the source of the UV light, we are bumping the nanobelt with light that has energy below the band-gap (3.60 eV). This way, we are excluding the formation of electron-hole pairs in the conduction/valence bands, that is why holes in the process of oxygen $O_2^-{}_{(ads)}$ desorption (**eq.(3)**) has to be thermally promoted from the acceptor energy level $E_A$, in the absence of UV-light, given rise to PPC.

## 6. Conclusions

In this work we have studied the photoconductivity (PC) of $SnO_2$ nanobelts as function of temperature and atmosphere (air, helium or vacuum), as well, low temperature photo-luminescence (PL) experiments.

In the PL experiment the optical transitions were studied in the visible-infrared region (VI-IR) and in the ultraviolet (UV) region. Peaks at the VI-IR region are associated to transitions from the oxygen-vacancy states to three donor levels, $E_{D1}$ = (60-80) meV, $E_{D2}$ = 130 meV and $E_{D3}$ = 210 meV, and peaks in the UV region to transitions from a donor level or from the conduction band to an acceptor level, $E_A$ = 230 meV. Not surprisingly the donor-level $E_{D1}$ were determined simultaneously from VI-IR and UV transitions, reinforcing the PL results presented.

In the PC measurements were observed that the maximum UV-induced photocurrent $(I_n)_{máx}$ increases with a decreasing in temperature, or in oxygen concentration, in the atmosphere around the sample. A persistent photoconductivity (PPC) effect was also observed, lasting for hours, even at temperatures as high as 400 K.

The kinetics of the PPC decay curves was found to be well explained by the stretched-exponential function, with an increase in the decay times as temperature decreases. Beside that a reduction of the available oxygen in the atmosphere around the sample (vacuum or helium) also modify drastically the PPC decay time.

The photoconductivity and the PPC results were explained in terms adsorption and desorption of molecular oxygen at the surface of the $SnO_2$ nanobelts, which promotes the creation of a band bending at the nanobelts surface, and consequentially spatial charge separation and a decrease in the free carriers' recombination time. The temperature dependence of the PPC decay time was found to exhibit activation energy around 230 meV that corresponds to the energy necessary for thermal ionization of free holes from acceptor levels to the valence band, in agreement with the photo-luminescence results presented. So, holes from ionized from acceptors levels are the responsible for the PPC effect in $SnO_2$, when we are using a light-source with energy below band-gap, that prohibit band to band transitions.

We claim that the experiments carried out in $SnO_2$ can be extrapolated for other metal-oxide semiconductors, especially for ZnO nanowires/nanobelts since both materials are naturally n-type semiconductors, having similar crystal structure and band gaps. Within the results presented the authors wants to exclude the common sense of the literature, that the oxygen vacancies are the striking forces on metal-oxide nanostructures in all aspects, in this case, it is not!

## Acknowledgments

The authors acknowledge the financial support of CNPq, CAPES and FAPEMIG, Brazilian funding agencies, and the technical support of the Microscopy Center of the Universidade Federal de Minas Gerais.